\documentclass[reprint,aps,superscriptaddress,prb,noshowpacs]{revtex4-1}%Use to see how Physical Review B looks

\usepackage{amsmath}
\usepackage{MnSymbol}
\usepackage{graphicx}% Include figure files
\usepackage{dcolumn}% Align table columns on decimal point
\usepackage{bm}% bold math
\usepackage{epstopdf}
\usepackage{color}
\newcommand{\be}{\begin{equation}}
\newcommand{\ee}{\end{equation}}

\newcommand{\ba}{\begin{eqnarray}}
\newcommand{\ea}{\end{eqnarray}}

\begin{document}
%\preprint{APS/123-QED}

%\title{Multiple spin correlations and the time evolution of a magic echo}
\title{Multispin correlations and pseudo-thermalization of the transient density matrix in solid-state NMR: free induction decay and magic echo}
\author{Steven W. Morgan}
\affiliation{%
Department of Physics, Brooklyn College, 2900 Bedford Avenue, Brooklyn,
NY 11210, USA}%
\author{Vadim Oganesyan}
\affiliation{Department of Engineering Science and Physics,
College of Staten Island, CUNY, 2800 Victory Blvd., Staten Island, NY 10314, USA}
\affiliation{The Graduate Center, CUNY, 365 5th Ave., New York, NY 10016}
\author{Gregory S. Boutis}%
\affiliation{%
Department of Physics, Brooklyn College, 2900 Bedford Avenue, Brooklyn,
NY 11210, USA}%
\affiliation{The Graduate Center, CUNY, 365 5th Ave., New York, NY 10016}
\date{\today}
\begin{abstract}
Quantum unitary evolution typically leads to
%equilibration (and
thermalization
%)
of generic interacting many-body systems.  There are very few known general methods for reversing this process, and we focus on the magic echo, a radio-frequency pulse sequence known to approximately ``rewind'' the time evolution of dipolar coupled homonuclear spin systems in a large magnetic field.  By combining analytic, numerical, and experimental results we systematically investigate factors leading to the degradation of magic echoes, as observed in reduced revival of mean transverse magnetization.
%polarizaiton, identifying finite pulse width effects as a dominant factor.
Going beyond the conventional analysis based on mean magnetization we use a phase encoding technique to measure the growth of spin correlations in the density matrix at different points in time following magic echoes of varied
%ous
durations and compare the results to those obtained during a free induction decay (FID). While considerable differences are %observed
documented at short times, the long-time behavior of the density matrix appears to be remarkably universal among the types of initial states considered -- simple low order multispin correlations are observed to decay exponentially at the same rate, seeding the onset of increasingly complex high order correlations. This manifestly athermal process is constrained by conservation of the second moment of the spectrum of the density matrix and proceeds indefinitely, assuming unitary dynamics.
\end{abstract}

\pacs{Valid PACS appear here}% PACS, the Physics and Astronomy

\maketitle
\section{\label{sec:intro}Introduction}
Nuclei in solids often enjoy relatively weak coupling to the environmental, non-magnetic degrees of freedom, marked by exceedingly long thermalization times, e.g. $T_1$ that can reach several weeks \cite{gatzke1993}.  Under these conditions coherent quantum many-body dynamics whose intrinsic time scales are typically some microseconds can %I put some microseconds because the Xe case given above has a T_2 of ~several hundred us.
be studied and manipulated with relative ease.  Much of the progress in elucidating local physics and chemistry of diverse substances
%materials 
over the past six decades has been achieved by matching detailed solutions of few-body dynamics with controlled experimental studies whereby through a combination of DC and RF fields desired many-body evolution can be realized to a high degree of precision.  The generation of so called ``echoes'' \cite{hahn1950} is a particularly central tool in the NMR arsenal. All echo schemes strive to pattern the finite time quantum evolution operator in such a way as to render it equal to the identity operator, thereby returning the many-body system
%ensemble
to its initial state. Most famously, the Hahn echo \cite{hahn1950} recovers the transverse magnetization dephased by local field inhomogeneities (and chemical shifts) with a simple reflection about an axis in the plane of spin precession. Remarkably, such single-spin corrective time-reversal action can be generalized with considerable success even to cases where the effective chemical shifts are time dependent (but mutually uncorrelated) \cite{uhrig2007}. Not surprisingly, time-reversing \emph{correlated} multispin dynamics is more subtle as the effective Hilbert space is no longer bounded.  To our knowledge there is no general systematic protocol for achieving this -- along with Maxwell's demon such general Loschmidt echoes \cite{gorin2006,zangara2012} were believed to be purely of conceptual significance. Yet, for dipolar spins subjected to strong DC fields an explicit solution to this problem, the so called ``magic echo,'' was found \cite{rhim71}, and has since served as the foundation for NMR studies of correlated spin motion, including measurements of imaging \cite{matsui1991,hafner1996}, spectroscopy \cite{takegoshi1985}, spin-diffusion \cite{zhang1998,boutis2004}, to investigate nonlinear dynamics in highly magnetized liquids \cite{hayden2007}, and, as in this work, for phase encoding and observing multispin correlations in non-commuting bases \cite{cho2005}.

%This
Our work pursues two different lines of inquiry which lead to a complementary characterization of the magic echo protocol. First, the experimentally observed efficacy of the magic echo, i.e. the relative amplitudes of the net refocused magnetization acquired using different magic echo conditions, are compared with
%predictions of
average Hamiltonian theory extended to second order, and also to results of numerical simulations.  Here, we specifically focus on the relative importance of the finite width of the RF pulses vs. the total RF power applied
% as
%These results identify finite pulse duration as
%the dominant source of degradation in efficacy
\footnote{The importance of accounting for finite pulse widths has been appreciated in numerous other contexts in NMR, see e.g. recent work on the persistence of coherence in long pulse trains\cite{li,dong}}. Second, the time dependence of the full density matrix  is characterized by measuring the onset and spreading of multispin correlations (also known as ``spin-counting'' spectroscopy) both during the usual free induction decay (FID, i.e. after a $\pi/2$ pulse from equilibrium) and following magic echoes with a few different, relatively long refocusing times. Interestingly, while these initial states are distinct and therefore evolve differently at short times their late time evolutions are apparently very similar, in the sense that certain rates of growth and decay are quantitatively indistinguishable. In the balance of the Introduction we summarize the basic physics of the magic echo and spin counting technique to help an impatient reader skip the following two Sections where a detailed discussion of notation and the theoretical framework (Section II), as well as experimental and numerical methodology (Section III), are given.  Results are presented and compared with theoretical expectations in Section IV with conclusions and discussion of future directions of research relegated to Section V.

The magic echo involves two key ingredients. First, the application of a strong DC field, $B_0$, along the $\hat{z}$ direction simplifies the interaction between two nuclei to
\begin{eqnarray}\label{eq:secdh}
\mathcal{H}_{D} &=& \frac{\gamma^2\hbar^2}{2r_{jk}^3}\left(1-3\cos^2\theta\right)\left(2I^0_j I^0_k-\frac{1}{2}(I^+_jI^-_k+I^-_jI^+_k)\right) \nonumber \\ &\equiv& D_{jk}[2I^0_j I^0_k-\frac{1}{2}(I^+_jI^-_k+I^-_jI^+_k)],%I'm actually not sure how much I like having the equations here; to me it would say just as much to put this later and just say you reverse the sign of the Hamiltonian during the spin locks.
\end{eqnarray}
where $\theta$ is the angle between the internuclear vector $\bm{r}_{jk}$ and the applied magnetic field which is oriented along the $z$-axis \cite{abragambook}, $\gamma$ is the gyromagnetic ratio of a nucleus, and $I^0, I^{\pm}$ are canonical spin operators. To set the notation for the remainder of the discussion the single spin (Larmor) precession frequency corresponding to the DC field is $\omega_{\rm L}=\gamma B_0$, while the internal dipolar fields can be associated with ``dipolar frequency'' $\omega_{\rm D}=\gamma^2/r^3$ ($r$ refers to the smallest internuclear distance). Typically $\omega_{\rm D}\sim~10^4~\rm{Hz}$, while $\omega_{\rm L}\sim~10^8~{\rm Hz}$ so the rapidly oscillating terms omitted from Eq. 1 are effectively suppressed by a small factor $\omega_{\rm D}/\omega_{\rm L}$. Also, in what follows we will always work in the frame rotating at  $\omega_{\rm L}$ (with respect to the lab frame).  Second, continuous application of a strong resonant field (i.e. at $\omega_{\rm L}$), $H_1$, in the $x-y$ plane ``locks'' the magnetization as the spins execute approximate Rabi oscillations at a frequency $\omega_1 = \gamma H_{1}$. In this Rabi frame of reference, co-rotating with the RF field, the dipolar interaction can be further reduced, approximately (i.e. assuming $\omega_D/\omega_1\to0$), to
\be
-\frac{D_{jk}}{2}[2\tilde{I}^0_j \tilde{I}^0_k-\frac{1}{2}(\tilde{I}^+_j\tilde{I}^-_k+\tilde{I}^-_j\tilde{I}^+_k)],
\ee
where $\tilde{I}^0,\tilde{I}^\pm$ refer to the quantization axis in the second rotating frame \cite{slichterbook}, i.e. if $H_1$ is along $\hat{x}$ then $\tilde{I}^0=I^x$. Most importantly, the relative minus sign of the $(0,0)$ and $(\pm,\mp)$ terms between Eqs. 1 and 2
%in the original dipolar kernel
translates into an overall sign reversal of spin interactions in the presence of the resonant field. After a few relatively minor additional steps the entire propagator during the application of the RF field (often referred to as an ``RF burst") becomes
\be
U_1=e^{+i \mathcal{H}_{D} t/(2\hbar)},
\ee
thus allowing the effective reversal (also referred to as ``refocusing'' or ``time-suspension'') of the entire many-body trajectory, which will typically consist of evolution under $U_{\rm D}\equiv e^{- i \mathcal{H}_{\rm D} t/\hbar}$ as well as $U_1$.

The idealized discussion above relies on a number of assumptions, of which the most tenuous is the availability of a very strong and precisely applied RF field. Quite generally, the efficacy of magic echoes is expected to degrade when longer trajectories are reversed either due to corrections beyond the zeroth order approximation in $\omega_{\rm D}/\omega_1$ (in the so-called average Hamiltonian theory)\cite{haeberlen1968,boutis2003,baudin2012}  or pulse artifacts (e.g. ring-down effects).  One generally expects the former (latter)  limitation to dominate at small (large) RF field strength, albeit for very different reasons (e.g. pulse artifacts are of purely experimental origin).
%, while the latter should become amplifed at strong RF power.
Surprisingly, very little is known about the degradation of magic echoes in the limit of long time suspension,
%evolutions
e.g. $\gtrsim$ 1 ms, and/or intermediate RF fields.  Although it is known how to achieve even longer time suspensions using short multiple (hundreds or even thousands) magic echoes in sequence \cite{rhim71,boutis2003}, some applications, e.g. the study of spin diffusion\cite{zhang1998,boutis2004}, can  benefit from the availability of a long duration single magic echo.
%, which suspends the dipolar evolution (including spin diffusion) during the application of field gradients.
%relies on the application of a very strong magnetic field gradient and therefore
 Understanding and eventually negating the factors leading to the degradation of long magic echoes, especially under realistic conditions of fixed total RF power applied to the sample, is a long term goal.%one of the goals of this work.
%relies on the application of a very strong magnetic field gradient to encode a grating in the magnetization.  Because of gradient limitations, this may require relatively long magic echo times.

While revival of the readily observable (uniform) magnetization is customarily taken as a proxy for the quality of the echo, we expect, at least in principle, the entire density matrix to be returned to near its initial (a $\pi/2$-rotated thermal) value $\rho(t=0)\sim{\bf 1}- \sum_j I^x_j$
%(as obtained by applying to the equilibrium magnetization a $\pi/2$ pulse that rotates $z\to x$)
. In the remainder of this work, we omit the identity term from $\rho(t=0)$, as is customary in NMR at infinite temperature. What can be said about the structure of the density matrix? The density matrix of nuclear spins in a rigid lattice during a conventional FID is described by the evolution of single-spin, single-quantum coherence (i.e. $\sum I_j^x$ term) into single-quantum, multiple spin correlations (i.e. terms involving $I_j^x I_k^z ...I_l^z$) . As these multispin correlations are not readily observable, the density matrix may appear to have thermalized but this is forbidden, strictly speaking, by the exact conservation of its spectrum under unitary evolution. As we will explain below, the experiment used to measure multispin correlations in fact accesses the second moment of the spectrum of the density matrix (the so called purity) and details the process of pseudo-thermalization, by which athermal correlations propagate (or spread) away from simple and readily observable parts of the density matrix to complex higher order correlations.

The early onset of these multispin correlations in a crystal
%can be measured, however, as demonstrated
has been measured in recent experiments \cite{cho2005} using a previously developed phase encoding technique. The basic trick essentially amounts to introducing steps to rotate the spin axis by a controlled amount thereby converting single-quantum operators into a linear superposition of operators of different coherence order, which can be distinguished. We extend this sort of study of the ``anatomy'' of the density matrix both to the late time regime for the usual FID but also, more interestingly, to post magic echo evolution. Direct comparison of short time dynamics reveals quantitative differences between the two types of the density matrices already at short times.
% we choose to focus on characterization at late times. While classification of all multi-spin operators \emph{and} precise connection between said classification and what the experiment measures is not straightforward the results of the experiment \emph{can} be expressed directly in terms of matrix elements of the density matrix and in turn, its spectrum.
Interestingly, and perhaps surprisingly, long time dynamics appears to be more universal. In addition to the
%assuming unitary time evolution one can ``derive'' a
``sum rule" mentioned above that guarantees that the growth of higher order correlations must come at the expense of lower order ones, we observe a high degree of similarity in the dynamics. For a sufficiently short-ranged Hamiltonian (which appears to include dipolar interations\cite{zobov2006}) both of these processes (decay and spread) appear to be simple exponential in time,
% processes
and we document the existence and correspondence of two multiple spin correlation decay rates in the same experiment \emph{and} across different initial states \cite{fine2004}.

\begin{figure}\begin{center}
\includegraphics[width=3.5in]{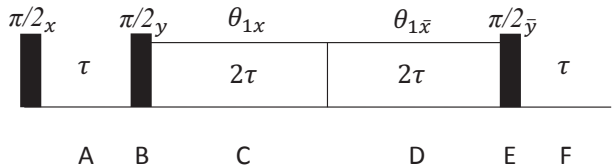}\caption{\label{fig:magicseq}The magic echo sequence used in this manuscript.  After an initial $\pi/2_{x}$ pulse, the spins evolve for a time $\tau$, followed by a sandwich consisting of two RF bursts along $x$ and $\bar{x}$, each of duration $2\tau$ between two $\pi/2$ pulses along $y$ and $\bar{y}$.  Because of the timing of this sequence, the echo appears at a time $\tau$ after the final $\pi/2$ pulse.  For the average Hamiltonian calculations the initial $\pi/2_{x}$ pulse is not considered.  The toggling frame Hamiltonians during time periods A-F are provided in Table \ref{tb:tfh}.}
\end{center}
\end{figure}
\section{\label{sec:sims}Theory of the magic echo and spin-counting spectroscopy}
In this Section we review and extend the average Hamiltonian theory of the magic echo, isolating finite pulse width effects from other (``burst") corrections contributing to an echo's demise. We also derive explicit expressions connecting the phase encoding experiment with density matrix elements and its purity.  All of this is available elsewhere, except, to the best of our knowledge, the first and second order average Hamiltonian calculations for the magic echo and the extension of the spin counting experiment to probe multiple spin correlations following a magic echo.%, and are reproduced here for uniformity and completeness of presentation.
\subsection{The Magic Echo}\label{sec:magicecho}
In a large, static magnetic field the spin dynamics in solid state NMR of spin-1/2 nuclei are dominated by the secular dipolar Hamiltonian $\mathcal{H}_{D}$.  To describe the effect of time dependent fields,
%the magic echo sequence
average Hamiltonian theory may be used \cite{haeberlen1968}.  In this formalism, the evolution of the density matrix is
\begin{eqnarray}
\label{eq:densitym}
\varrho(\tau_{c}) &=& U_d U_{RF}U_d\varrho(0)(U_d U_{RF}U_d)^{-1} \nonumber \\ &\equiv& U(\tau_c)\varrho(0)U^{-1}(\tau_c).
\end{eqnarray}
The propagators $U_{RF}$ and $U_d$ refer to evolution with and without the RF field, respectively, and $\tau_c$ is the time for a complete magic echo cycle.  The propagator $U_{RF}$
may be written by the Dyson series\begin{equation}\label{eq:URF}
U_{RF}(t) = \mathcal{T}\exp{(-i\int_{0}^{t}\mathcal{H}_{RF}(t_{1})dt_{1})}.
\end{equation}In the above expression $\mathcal{T}$ is the time ordering operator and $\mathcal{H}_{RF}$ is the RF Hamiltonian in the rotating reference frame (which is explicitly time-dependent).  The object of average Hamiltonian theory is to impart a well-defined time dependence on the internal interaction, which is otherwise time independent, via a sequence of experimentally controlled RF pulses. Over a cycle of RF pulses of duration $\tau_{c}$, the propagator $U$ may be written in terms of the Magnus expansion\cite{Maricq1982}
%, i.e. as:
\begin{equation}\label{eq:Umagnus}
U(\tau_{c}) = e^{-i(\bar{\mathcal{H}}^{0}+\bar{\mathcal{H}}^{1}+\ldots)\tau_{c}}.
\end{equation}
The zeroth, first, and second order terms in the Magnus expansion are
\begin{widetext}
\begin{eqnarray}
\label{eq:zerothavgh}
\bar{\mathcal{H}^{0}} & =&  \frac{1}{\tau_{c}}\int_{0}^{\tau_{c}}\tilde{\mathcal{H}}(t_{1})dt_{1},
\\
\label{eq:firstavgh}
\bar{\mathcal{H}^{1}} & = & -\frac{i}{2\tau_{c}}\int_{0}^{\tau_{c}}dt_{2}\int_{0}^{t_{2}}\left[\tilde{\mathcal{H}}(t_{2}),\tilde{\mathcal{H}}(t_{1})\right]dt_{1},
\\
\label{eq:secondavgh}
\bar{\mathcal{H}^{2}} &=&  -\frac{1}{6\tau_{c}}\int_{0}^{\tau_{c}}dt_{3}\int_{0}^{t_{3}}dt_{2}\int_{0}^{t_{2}}\left(\left[\tilde{\mathcal{H}}(t_{3}),\left[\tilde{\mathcal{H}}(t_{2}),\tilde{\mathcal{H}}(t_{1})\right]\right]+  \left[\tilde{\mathcal{H}}(t_{1}),\left[\tilde{\mathcal{H}}(t_{2}),\tilde{\mathcal{H}}(t_{3})\right]\right] \right)dt_{1},
\end{eqnarray}
\end{widetext}
where $\tau_c=6 \tau+ 2 t_{\pi/2}$ for the magic echo (see Fig. \ref{fig:magicseq}). Further analysis is facilitated by defining the so-called ``toggling frame Hamiltonian'' which in the absence of an RF field is rotating at the Larmor frequency $\omega_L$ about $\hat{z}$ and is otherwise rotating about the RF field at $\omega_{1}$, with $\tau\gg t_{\pi/2}\equiv\frac{\pi}{2\omega_{1}}$, the duration of the $\pi/2$ pulses.
We list the toggling frame Hamiltonians $\tilde{\mathcal{H}}$ for the magic echo sequence shown in Fig. \ref{fig:magicseq} in Table \ref{tb:tfh}.
% In this formalism, the toggling frame Hamiltonian $\tilde{\mathcal{H}}$ is given by
%\begin{equation}
%\label{tfrh}
%\tilde{\mathcal{H}} = U^{-1}_{RF}(t)\mathcal{H}U_{RF}(t),
%\end{equation}
 %where $\mathcal{H}$ is the Hamiltonian that is not explicitly time-dependent (for the case here of dipolar coupled spins it is the secular dipolar Hamiltonian, Eq. \ref{eq:secdh}).
\begin{table*}
\caption{
\label{tb:tfh}The toggling frame Hamiltonians for the magic echo pulse sequence depicted in Fig. \ref{fig:magicseq}.  Here $\theta_{1} = 2\tau\omega_{1}$ is the flip angle of the first $x$ RF burst as seen in Fig. \ref{fig:magicseq} (time period C).}
\begin{ruledtabular}
\begin{tabular}{lll}
Time period & Duration & Toggling frame Hamiltonian \\ \hline
A & $\tau$& $D_{12}(3I_{1}^z I_{2}^z - \bm{I_{1}}\cdot\bm{I_{2}})$ \\
B & $t_{\pi/2}\equiv\pi/(2 \omega_1)$ & $3D_{12}[I_{1}^zI_{2}^z\cos^{2}{(\omega_{1}t)} + \left(I_{1}^z I_{2}^x+I_{1}^x I_{2}^z \right)\cos{(\omega_{1}t)}\sin{(\omega_{1}t)}+$ $I_{1}^x I_{2}^x \sin^{2}{(\omega_{1}t)}]-D_{12}\bm{I_{1}}\cdot\bm{I_{2}}$ \\
C & $2\tau$ & $3D_{12}[I_{1}^x I_{2}^x\cos^{2}{(\omega_{1}t)}-(I_{1}^xI_{2}^y+I_{1}^y I_{2}^x)\cos{(\omega_{1}t)}\sin{(\omega_{1}t)}+$  $I_{1}^y I_{2}^y\sin^{2}{(\omega_{1}t)}]-D_{12}\bm{I_{1}}\cdot\bm{I_{2}}$ \\
D & $2\tau$ & $3D_{12}[I_{1}^x I_{2}^x\cos^{2}{(-\omega_{1}t+\theta_{1})}+(I_{1}^x I_{2}^y+I_{1}^y I_{2}^x)\sin{(-\omega_{1}t+\theta_{1})}\cos{(-\omega_{1}t+\theta_{1})}+$  $I_{1}^y I_{2}^y \sin^{2}{(-\omega_{1}t+\theta_{1})}]$ \\& & $- D_{12}\bm{I_{1}}\cdot\bm{I_{2}}$ \\
E & $t_{\pi/2}\equiv\pi/(2 \omega_1)$& $3D_{12}[I_{1}^z I_{2}^z\sin^{2}{(\omega_{1}t)} + (I_{1}^z I_{2}^x+I_{1}^x I_{2}^z)\sin{(\omega_{1}t)}\cos{(\omega_{1}t)}+$  $I_{1}^x I_{2}^x\cos^{2}{(\omega_{1}t)}]-D_{12}\bm{I_{1}}\cdot\bm{I_{2}}$ \\
F & $\tau$ & $D_{12}(3I_{1}^z I_{2}^z - \bm{I_{1}}\cdot\bm{I_{2}})$ \\

\end{tabular}
\end{ruledtabular}
\end{table*}

In our analysis of the average Hamiltonian we only show interactions between pairs of spins.  Because of the commutators in Eqs. \ref{eq:firstavgh} and \ref{eq:secondavgh}, operators involving three or more spins generally appear in the higher-order terms--we neglect these terms for simplicity. The first three terms
% order term
in the Magnus expansion for the secular dipolar Hamiltonian in the magic sandwich shown in Fig. \ref{fig:magicseq} are given by (note that Ref. \onlinecite{boutis2003} does this zeroth order average Hamiltonian calculation for a slightly different variation of the magic echo sequence):
\begin{widetext}
\begin{eqnarray}
%\bar{\mathcal{H}_{D}^{0}}  = &-&D_{12}\bm{I_{1}}\cdot\bm{I_{2}} + D_{12}I_1^{z}I_2^{z}\left(1+\frac{\pi/2\omega_{1}}{6\tau+\pi/\omega_{1}}\right)
 %+ \frac{3D_{12}}{6\tau + \pi/\omega_{1}}I_1^{x}I_2^{x}\left[2\left(\tau+\frac{\sin(4\tau\omega_{1})}{4\omega_{1}}\right)+\frac{\pi}{2\omega_{1}} \right] \nonumber \\ & + &\frac{3D_{12}}{6\tau+\pi/\omega_{1}}\left(I_1^{x}I_2^{y}+I_1^{y}I_2^{x}\right)\left[\frac{\sin^{2}(2\tau\omega_{1})}{4\omega_{1}}\right] \nonumber \\ &+& \frac{3D_{12}}{6\tau+\pi/\omega_{1}}\left[2 I_1^{y}I_2^{y}\left(\tau - \frac{\sin(4\tau\omega_{1})}{4\omega_{1}}\right) + \left(\frac{I_1^{z}I_2^{x}+I_1^{x}I_2^{z}}{\omega_{1}}\right)\right].\label{eq:zeroah}
%\\
\bar{\mathcal{H}_{D}^{0}}&=&\frac{D_{12} t_{\pi/2}}{\tau_c}\left[{\bf I_1}\cdot{\bf I_2}-3I_1^y I_2^y+\frac{6}{\pi}(I_1^z I_2^x+I_1^x I_2^z)\right]+\frac{3D_{12}}{2\tau_c \omega_1}\left[\sin (4 \omega_1 \tau) (I_1^x I_2^x- I_1^y I_2^y)+\frac{\sin^2(2\omega_1 \tau)}{2}(I_1^xI_2^y+I_1^yI_2^x)\right]
\end{eqnarray}
\begin{eqnarray}
\bar{\mathcal{H}_{D}^{1}}& =& -\frac{9D_{12}^{2}\sin^{2}{(2\tau\omega_{1})}\sin{(4\tau\omega_{1})}(I_1^{z}+I_2^{z})}{16\omega_{1}^{2}\tau_c}
\label{eq:avh1o}
\end{eqnarray}
\begin{eqnarray}
\label{eq:secondah}
\bar{\mathcal{H}_{D}^{2}} =&-&\frac{9D_{12} t_{\pi/2}}{16\pi \tau_c}[\frac{D_{12}^{2}}{\omega_{1}^{2}} ( 3\tau \omega_{1}  + \frac{\sin(4 \tau \omega_{1})}{2}) (4I_1^{z}I_2^{z}-2I_1^{x}I_2^{x}-2I_1^{y}I_2^{y}) \nonumber \\
&+&\frac{D_{12}^{2}}{\omega_{1}^{2}}(-\frac{1}{16}+\tau^{2}\omega_{1}^{2}+ \frac{\cos(4\tau\omega_{1})}{8}-\frac{\cos(8\tau\omega_{1})}{16}  +  \tau\omega_{1}\sin(4\tau\omega_{1}))(4I_1^{x}I_2^{z}+ 4I_1^{z}I_2^{x}) \nonumber \\
&+& \frac{D_{12}^{2}}{\omega_{1}^{2}}(-\frac{3\pi}{16}- 3\tau\omega_{1} +\frac{\pi\cos(4\tau\omega_{1})}{4} - \frac{\pi\cos(8\tau\omega_{1})}{16}- \frac{\tau\omega_{1}\sin(4\tau\omega_{1})}{2})(2I_1^{x}I_2^{x}-2I_1^{y}I_2^{y})].
\end{eqnarray}
 \end{widetext}
In the limit of infinite $\omega_{1}$ the above expressions vanish, as expected. With $\bar{\mathcal{H}}_{D}^{n}$=0 ($n=0,1,2 \ldots$), one refocuses the many-body spin dynamics resulting in the observation of a magic echo.
% (i.e. $D_{12}/\omega_{1} \rightarrow 0$),
%the zeroth order term is observed to vanish \cite{rhim71}. Thus, to zeroth order, with $\bar{\mathcal{H}_{D}^{0}}$=0 one refocuses the many-body spin dynamics resulting in the observation of a magic echo.
%Using Eq. \ref{eq:firstavgh} and the toggling frame Hamiltonians given in Table \ref{tb:tfh} it may be shown that the first order term, $\bar{{\mathcal{H}}^{1}_{D}}$, is given by
%\begin{equation}
%\bar{\mathcal{H}_{D}^{1}} = -\frac{9D_{12}^{2}\sin^{2}{(2\tau\omega_{1})}\sin{(4\tau\omega_{1})}(I_1^{z}+I_2^{z})}{16\omega_{1}^{2}(6\tau + \pi/\omega_{1})}
%\label{eq:avh1o}
%\end{equation}
%and the second order term in the Magnus expansion is
%\begin{widetext}
%\begin{eqnarray}
%\label{eq:secondah}
%\bar{\mathcal{H}_{D}^{2}} &=&\frac{-9}{32(6\tau + \pi/\omega_{1})}[\frac{D_{12}^{3}}{\omega_{1}^{3}} \left[ 3\tau \omega_{1}  + \frac{\sin(4 \tau \omega_{1})}{2}\right] \left(4I_1^{z}I_2^{z}-2I_1^{x}I_2^{x}-2I_1^{y}I_2^{y}\right) \nonumber \\ &+&
%\frac{D_{12}^{3}}{\omega_{1}^{3}}[-\frac{1}{16}+\tau^{2}\omega_{1}^{2}+ \frac{\cos(4\tau\omega_{1})}{8}-\frac{\cos(8\tau\omega_{1})}{16}  +  \tau\omega_{1}\sin(4\tau\omega_{1})]\left(4I_1^{x}I_2^{z}+ 4I_1^{z}I_2^{x}\right) \nonumber \\ &+& \frac{D_{12}^{3}}{\omega_{1}^{3}}[-\frac{3\pi}{16}- 3\tau\omega_{1} +\frac{\pi\cos(4\tau\omega_{1})}{4} - \frac{\pi\cos(8\tau\omega_{1})}{16}- \frac{\tau\omega_{1}\sin(4\tau\omega_{1})}{2}](2I_1^{x}I_2^{x}-2I_1^{y}I_2^{y})].
%\end{eqnarray}
%\end{widetext}
The zeroth order term term contains imperfections from both pulse width effects and finite RF during the burst. The first order term, $\bar{\mathcal{H}_{D}^{1}}$, results from the commutator between the toggling frame Hamiltonians during times C and D in Fig. \ref{fig:magicseq} and is, therefore, free of pulse width effects.  This term vanishes if the durations of the RF bursts are such that $2 \omega_{1} \tau$ is a multiple of $\pi$ and/or if $D_{12}/\omega_{1} \rightarrow 0$.  Even if $\bar{\mathcal{H}^{1}_{D}} \neq 0$, 
%$\bar{\mathcal{H}^{1}_{D}}$ 
it cannot degrade the magnitude of transverse magnetization -- it involves only $I_{iz}$ operators and only produces a rotation about the $z$-axis. The second order term in the Magnus expansion, $\bar{\mathcal{H}_{D}^{2}}$, arises \emph{only} from commutators including time periods B and/or E (the $\pi/2$ pulses of the magic sandwich) -- this term vanishes for infinitely narrow ($\delta$-function) $\pi/2$ pulses in the magic sandwich.  Importantly, for finite $t_{\pi/2}$ it includes terms which increase with echo time $\tau$. Thus, to second order finite RF pulses will induce imperfections in the refocusing offered by the magic sandwich with increasing $\tau$ which may be evident in both the multiple spin correlations and in the single-quantum signal from a magic echo.
\subsection{Multiple Spin Correlations}
\begin{figure*}\begin{center}
\includegraphics[width=7in]{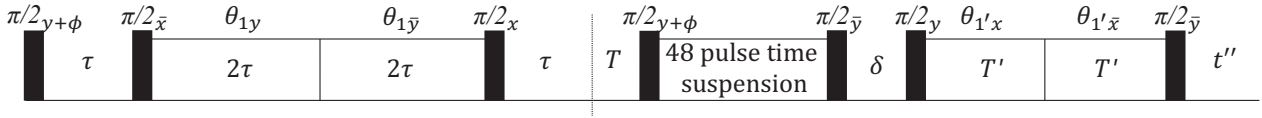}\caption{\label{fig:pulseseq}Pulse sequence used to observe multiple spin correlations following a magic echo.  For the observation of multiple spin correlations following the FID, the first magic sandwich is removed as are the delays $\tau$ before and after it.  The time suspension sequence is used to suppress artifacts and suspend evolution under the dipolar interaction \cite{cory1990}.  The delays $T'$ and $\delta$ are adjusted so that the refocusing of the dipolar interaction peaks after instrumental ringdown following the end of the final magic sandwich.  The flip angle $\theta_{1'} = \omega_{1}T'$.}
\end{center}
\end{figure*}

%To study the multiple spin correlations in the magic echo we use a pulse sequence (shown in Fig. \ref{fig:pulseseq}) similar to one used to measure the multiple spin correlations following the magic echoes.  This sequence is similar to one used previously \cite{cho2005,sanchez2009} to measure multiple spin correlations in the FID.  The notable addition is the first magic echo (including the evolution $\tau$ before and after the first magic sandwich).  We omit this portion to measure the multiple spin correlations following the FID for comparison with those we measure following the magic echoes, in which case the sequence is nearly identical to that used in Ref. \onlinecite{cho2005}.
When NMR signals are detected by inductive coupling to a coil the measured signal at time $t$ is given by
\begin{equation}
\label{eq:sigmeas}
S(t) \sim \textrm{Tr}[\varrho(t) I^{+}].
\end{equation}
Only single spin, single quantum terms survive the trace. The growth of multiple spin single quantum coherence can be observed in the short time expansion of the density matrix
%To describe the measurements of the multiple spin correlations using the pulse sequence in Fig. \ref{fig:pulseseq}, we start with a $\pi/2$ pulse which rotates the initial magnetization into the transverse plane.  When observing the multiple spin correlations during the FID the magnetization evolves during time $T$ under the dipolar Hamiltonian, Eq. \ref{eq:secdh}.  In the case of spin correlations following the magic echo, the Hamiltonian is a combination of the dipolar evolution starting at the beginning of time period $T$ and the sum of all the terms of the Magnus expansion of the magic echo.  The time evolution of the density matrix under the resulting Hamiltonian $H$ is
%\begin{equation}
%\label{eq:varrrhot}
%\varrho(t) = e^{-iHt/\hbar}\varrho(0)e^{iHt/\hbar}
%\end{equation}and may be expanded as a series:
\begin{equation}
\label{eq:varrhoexp}
\varrho(t) = \varrho(0) + \frac{i}{\hbar}t\left[\varrho(0),H\right] - \frac{t^{2}}{2\hbar^{2}}\left[\left[\varrho(0),H\right],H\right]+\ldots,
\end{equation}
where
\begin{equation}\label{eq:initialcond}
\varrho(0)\sim -\sum_{j}I_j^{x}
\end{equation}
after a ${\pi/2}_{y}$ pulse (i.e. $\phi = 0$ in Fig. \ref{fig:pulseseq}).  Using Eq. \ref{eq:secdh} for the Hamiltonian the evolution leads to
\begin{widetext}\begin{eqnarray}\label{eq:FIDzb}
\varrho_{F}(t) = &-&\frac{1}{2}\sum_{j}\left(I_j^{+}+I_j^{-}\right) + 3it\sum_{jk}D_{jk}\left(-I_j^{0}I_k^{+}+I_j^{0}I_k^{-}\right) \nonumber \\&-&  3 t^{2}\sum_{jkl}D_{jk}D_{kl}[I_l^{0}I_j^{0}I_k^{+}+I_l^{0}I_j^{0}I_k^{-}+I_l^{0}I_j^{+}I_k^{0}/2+I_l^{0}I_j^{-}I_k^{0}/2+ I_l^{+}I_k^{-}I_j^{+}/4-I_l^{+}I_k^{+}I_j^{-}/4 \nonumber \\ &-&I_l^{-}I_k^{-}I_j^{+}/4+I_l^{-}I_k^{+}I_j^{-}/4]+\ldots,
\end{eqnarray}\end{widetext}
with $t$ within the time interval labelled $T$ in Fig. \ref{fig:pulseseq}.  The insight into how to begin resolving the different terms in the density matrix comes from simply rewriting it in another, non-commuting basis, e.g. polarized along $\hat{x}$
\begin{widetext}
\begin{eqnarray}\label{eq:FIDxb}\tilde{\varrho}_{F}(t) = &-&\sum_{j}\tilde{I}_j^{0} - \frac{3}{2}it\sum_{jk}D_{jk}(\tilde{I}^{+}_{j}\tilde{I}^{+}_{k}-\tilde{I}^{-}_{j}\tilde{I}^{-}_{k}) + \frac{3}{2}t^{2}\sum_{jkl}D_{jk}D_{kl}(\frac{3}{2}\tilde{I}^{+}_{j}\tilde{I}^{+}_{l}\tilde{I}^0_{k}-\frac{1}{2}\tilde{I}^{+}_{j}\tilde{I}^{-}_{l}\tilde{I}^0_{k} \nonumber \\&-&\frac{1}{2}\tilde{I}^{-}_{j}\tilde{I}^{+}_{l}\tilde{I}^0_{k}+\frac{3}{2}\tilde{I}^{-}_{j}\tilde{I}^{-}_{l}\tilde{I}^0_{k}-\tilde{I}^-_{l}\tilde{I}^+_{k}\tilde{I}^0_{j}-\tilde{I}^+_{l}\tilde{I}^-_{k}\tilde{I}^0_{j})+\ldots.\end{eqnarray}\end{widetext}
It is clear that upon basis rotation single quantum and zero quantum operators are rotated into each other, producing operators changing polarization along $x$ by arbitrary amounts. The growth of such observable $\hat{x}$-multiquantum operators can be taken as  a proxy for the onset of multispin correlations (in fact, the two are uniquely related).  All that is left is to ``count" coherence orders in the non-commuting basis, which is done by introducing phase-modulated rotations, as in Fig. 2 and Eq. 19.
In the experiment shown in Fig. \ref{fig:pulseseq} the transformation of basis is performed using the first, fourth, and fifth $\pi/2$ pulses.  To encode odd orders, the first pulse is phase shifted by $\pi/2$ from what is shown in the figure.  In the present work, with the magnetization in the $x$ basis, the phase $\phi$ is varied from $0$ to $4\pi$ with 120 total values of $\phi$ to encode up to the $\pm$30th order of spin correlations.  The zeroth order spin correlations have coherence order 0 in the $x$-basis, corresponding to the first term plus four of the six terms in the quadratic-time-dependent portion of Eq. \ref{eq:FIDxb} as well as those of coherence order zero from higher orders in time.  The higher-order terms are converted back into single-spin terms by a magic echo, i.e. the final magic echo of Fig. \ref{fig:pulseseq}.  For signal processing, one implements a Fourier transform of the resulting signal (with respect to $\phi$) to generate a spectrogram of the coherence orders.  Because of the mixing of coherence order $N$ among spin correlations with $N$ or more spins, the signals do not correspond directly with $N$-spin single-quantum coherence terms from Eq. \ref{eq:FIDzb} but instead represent the correlations among at least $N$ spins.
Defining a shorthand for the total moment $\vec{I}=\sum_j \vec{I}_j$ we can write the (inverse) propagator of the pulse sequence in Fig. \ref{fig:pulseseq} explicitly
\begin{widetext}
\begin{equation}\label{eq:prop}
[U_{\phi}]^{-1} = e^{-i\phi I^{z}}e^{i\frac{\pi}{2}I^{y}}e^{i\phi I^{z}}e^{iH\tau}e^{-2iH\tau}e^{iH\tau}e^{iHT}e^{-i\phi I^{z}}e^{i\frac{\pi}{2} I^{y}} e^{i\phi I^{z}}e^{-i\frac{\pi}{2}I^{y}} e^{i\delta H}e^{-iT'H}e^{iHt''}
\end{equation}
\end{widetext}
where $t''$ is the time after the last pulse and $H$ is the dipolar Hamiltonian.  This propagator explicitly assumes that the second magic sandwich perfectly refocuses the dynamics (this assumption is valid as this magic echo is relatively short).  The first magic echo will be absent if one is studying multispin correlations in the FID. With $t' = T' - t'' - \delta$ the peak of the echo is located at $t' = T$ while the propagator simplifies to:
\begin{equation}\label{eq:prop2}
[U_{\phi}]^{-1} = e^{-i\phi I^{z}}e^{i\frac{\pi}{2} I^{y}}e^{iHT}e^{i\phi I^{x}}e^{-iHt'}\end{equation}
and the observed signal is
\begin{eqnarray}S_{\phi} &=& \left<U_{\phi}^{-1}I^{z}U_\phi I^{x}\right>\nonumber \\&=& \left<e^{iHt'}e^{-i\phi I^{x}}e^{-iHT} I^{x} e^{iHT}e^{i\phi I^{x}}e^{-iHt'}I^{x}\right>  \nonumber \\&=&\left<m|e^{-i\phi I^{x}/2} I^{x}(T)e^{i\phi I^{x}/2}|n\right>\left<n|e^{i\phi I^{x}/2} I^{x}(t')e^{-i\phi I^{x}/2}|m\right>\nonumber \\ & = &\sum_{n,m} e^{i(n-m)\phi} [I^{x}(T)]_{mn}[I^{x}(t')]_{nm}\equiv\sum_n \tilde{S}_n e^{i n \phi}.\label{eq:signalphi}
\end{eqnarray}
Note that in addition to assuming a perfect magic echo this protocol also assumes perfect conservation of $I^z$, which is valid in the high field limit. Clearly,  $\sum_n \tilde{S}_n = \textrm{Tr}[\rho^2$] for $t'=T$.
\section{\label{sec:exp}Methodology}
\subsection{Experimental procedures and sample characterization}
\begin{figure}
\includegraphics[width=3.5in]{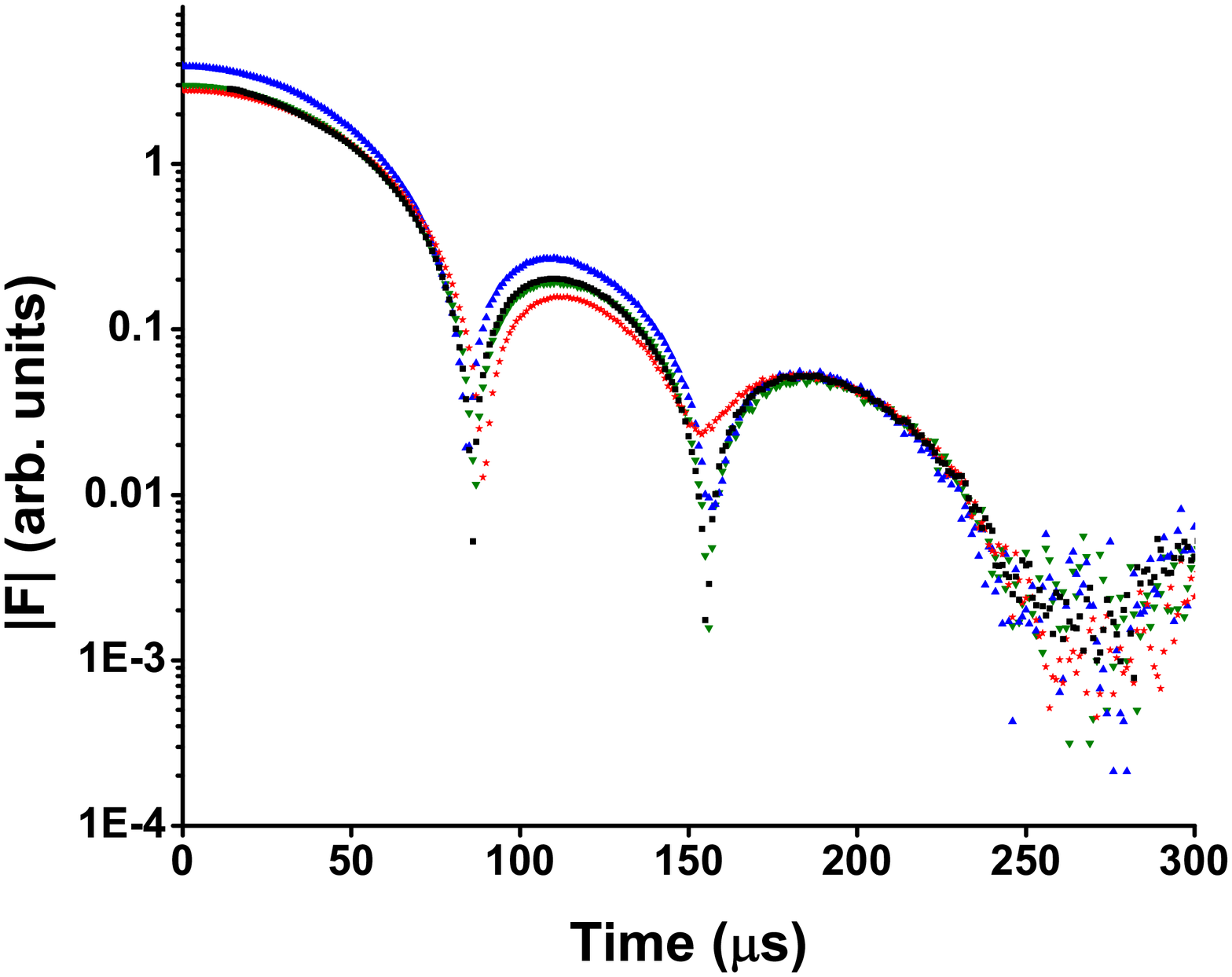}\caption{\label{fig:LTOL}(Color online) Experimental FID and magic echoes measured in adamantane.  To overlay the data, we have time-shifted and normalized each echo decay. Even though the initial decays vary significantly, they all approach the same long-time decays.  In this figure the points represented are: the FID ($\filledsquare$/black) and magic echoes at different $\tau$,$\tau$= $100~\mu$s  ($\filledstar$/red), $\tau$=$150~\mu$s ($\filledtriangleup$/blue), and $\tau$= $200~\mu$s ($\filledtriangledown$/green).}
\end{figure}

\begin{figure}
\includegraphics[width=3.5in]{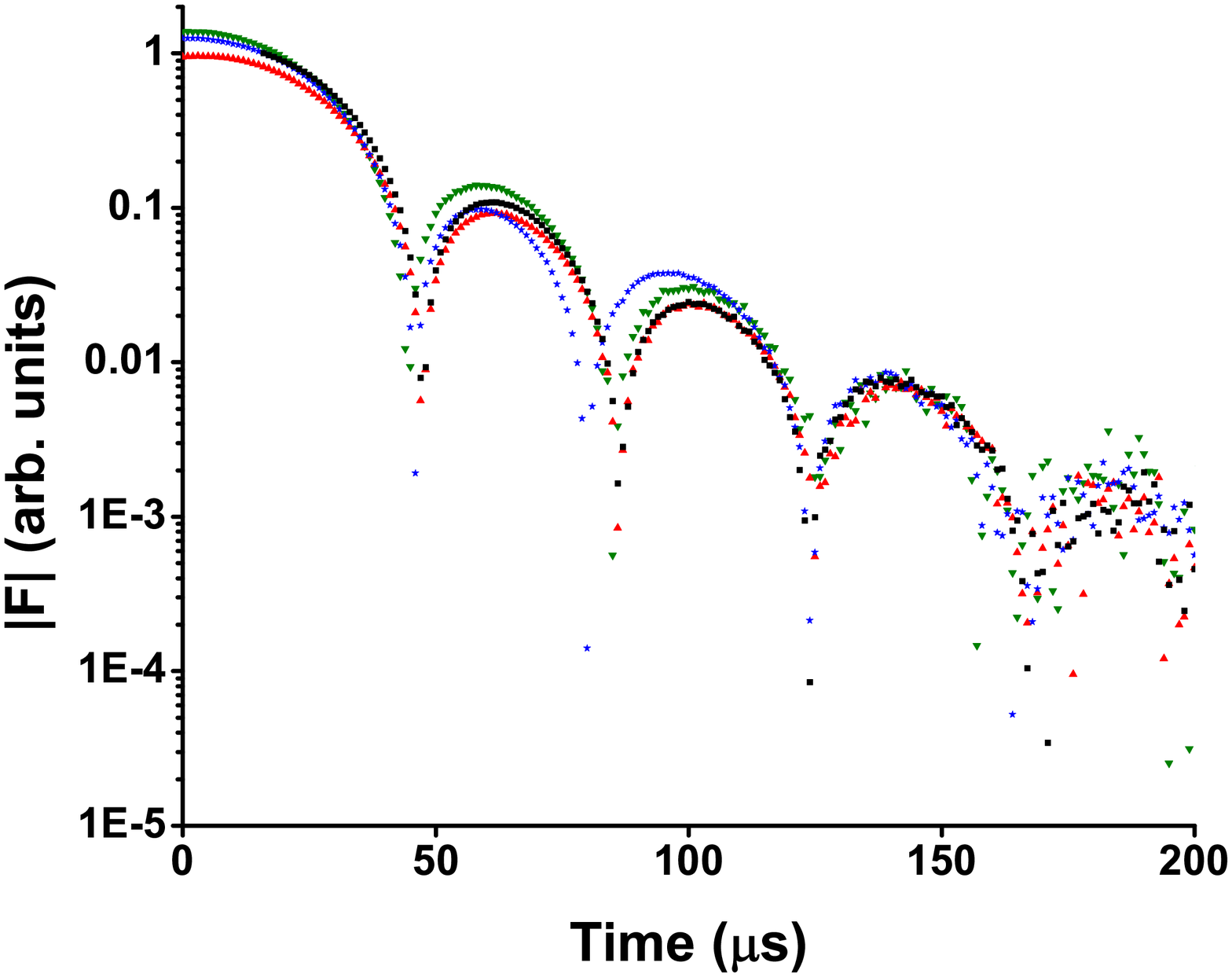}\caption{\label{fig:LTOLCaF2}(Color online) Experimental FID and magic echoes measured in calcium fluoride.  To overlay the data, we have time-shifted and normalized each echo decay. Even though the initial decays vary significantly, they all approach the same long-time decays.  In this figure the points represented are: the FID ($\filledsquare$/black) and magic echoes at different $\tau$, $\tau$=$50~\mu$s ($\filledstar$/red), $\tau$= $100~\mu$s ($\filledtriangleup$/blue), and $\tau$=$130~\mu$s ($\filledtriangledown$/green).}
\end{figure}

Experiments were performed in adamantane and calcium fluoride at room temperature.  Adamantane is a plastic crystal in which molecular tumbling averages out intramolecular interactions \cite{schnell2001}, leaving every proton to interact with each of the 16 protons in all of the 12 nearest neighbor molecules in addition to those further away.  Because of tumbling these molecules act as if all the protons in a molecule are at the center of that molecule.  The effective dipolar coupling $D_{\textrm{eff}}$ in adamantane is $\approx 4.7 \times 10^{4}$ rads/s \cite{schnell2001}, and is determined using a measurement of the linewidth. The calcium fluoride sample used in this study was a single crystal oriented such that the duration between the first two zero crossings in the FID (note that in Figs. \ref{fig:LTOL} and \ref{fig:LTOLCaF2} only the magnitude of the signal is plotted) is approximately $40~\mu$s, which lies between the values for the [110] and [111] axes along the Zeeman field.  $^{1}$H and $^{19}$F NMR experiments were performed at 179.445 and 168.824 MHz using a Tecmag Apollo spectrometer.

In Figs. \ref{fig:LTOL} and \ref{fig:LTOLCaF2} we show a series of magic echoes and FIDs in adamantane and calcium fluoride respectively.  The data are normalized and time-shifted to emphasize the overlap in the long-time portion of the decays \cite{morgan2008}.  In the case of the magic echoes, the initial portions of the signals vary with echo time but eventually decay and oscillate at the same rate.  We show the results $g$ and $\Omega$ of a fit of the long-time portion of the decays to
\begin{equation}
\label{eq:ltsqc} F(t) = F_0\exp{(-g t)}\sin{(\Omega t + \Phi)}\end{equation} in Table \ref{tb:ltfitsq} for adamantane and calcium fluoride.  This form for the long-time portion of the FID has a long history \cite{borckmans1968} and has recently been predicted using the notion of microscopic chaos \cite{fine2004,fine2005}. The sinusoidally modulated exponential characteristic has also been observed for the FID when measured out to six orders of magnitude \cite{meier2012}.  We note that
%although
our measurements for calcium fluoride do not correspond to those in Ref. \onlinecite{meier2012}, because our crystal is not exactly in the [110] orientation.
%Additionally, \cite{meier2012} showed that there are at least two modes in the long-time FIDs of CaF$_{2}$ in the [110] and [100] orientations.  This long-time universal regime is further probed by observing the the time evolution of multiple spin correlations in following experiments.

\begin{figure}
\includegraphics[width=3.5in]{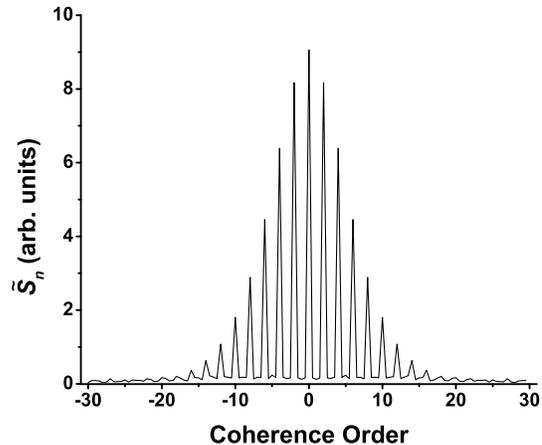}
\caption{\label{fig:COHOE}Representative even order experimental multiple spin correlations, obtained in adamanatane for magic echo time $\tau = 100~\mu$s and evolution time $T=240~\mu$s (see Fig. \ref{fig:pulseseq} for definitions of $\tau$ and $T$).  We performed 4 signal averages and incremented $\phi$ from $0$ to $4\pi$ by $\pi/30$ (120 phases $\phi$ were used).  The spectrogram shown in the figure is obtained by a Fourier transform of the echo peak amplitude variation with phase.  The growth and subsequent decay of coherence peak amplitudes for adamantane and calcium fluoride are plotted in Figs. \ref{fig:ltmqc} and \ref{fig:ltmqcCaF2}}.
\end{figure}

\begin{table*}
\caption{\label{tb:ltfitsq}Values of the parameters $\Omega$ and $g$ determined for the fit of the FID and magic echo signals to the equation $F_0\exp{(-g t)}\sin{(\Omega t + \Phi)}$ in adamantane and calcium fluoride.
 The parameters shown in the tables (and associated error bars) were determined from six independent fits to the data.
%The parameters are from six separate fits to data.
}\begin{ruledtabular}
\begin{tabular}{cccccc}
adamantane & $\Omega$ (rads/ms) & $g$ (ms$^{-1}$) & CaF$_{2}$ & $\Omega$ (rads/ms) & $g$ (ms$^{-1}$) \\
\hline
FID & $ 30 \pm 1$ & $32 \pm 4$ & FID & $71 \pm 1$ & $34 \pm 3$\\
$100~\mu$s magic echo & $31 \pm 1$ & $31 \pm 3$ & $50~\mu$s magic echo & $73 \pm 2$ & $37 \pm 2$\\
$150~\mu$s magic echo & $36 \pm 1$ & $33 \pm 2$ & $100~\mu$s magic echo & $74 \pm 2$ & $33 \pm 3$ \\
$200~\mu$s magic echo & $33 \pm 1$ & $34 \pm 3$ & $130~\mu$s magic echo & $75 \pm 3$ & $33 \pm 3$\\
\end{tabular}
\end{ruledtabular}
\end{table*}
Figure \ref{fig:COHOE} highlights a representative experimental spectrogram for spin correlations measured in adamantane with a magic echo time $\tau = 100~\mu$s and an evolution time $T = 240~\mu$s. In all experiments performed, we used 4 signal averages with a recycle delay of 5 (30) seconds in adamantane (CaF$_{2}$). In our experiments, $\delta$ was set to $10~\mu$s and $T' = T + 90 (70)~\mu$s in adamantane (CaF$_{2}$) to avoid probe ring down artifacts.  Our experimental pulse sequence also implemented a simple phase cycle to remove baseline artifacts and imbalance in the receiver channels \cite{hoult1975}. The time of the echo peak for the $\phi = 0$ phase encoding step was used when computing the spectrogram.
\subsection{Notes on simulations}
Numerical simulations were performed by integrating equations of motion for the expectation values of spins in the classical limit, with quantum spin operators replaced by ``classical spins,'' i.e. three component unit vectors obeying the appropriate Poisson bracket. These deterministic Bloch equations were integrated using a 4th order Runge-Kutta method with a fixed time step, $\Delta t$.  Optimally, this time step is chosen to be short enough to sample single spin precession at least $\sim$ 10 times per typical precession period but not too short to overburden the computation. Going over and performing the entire simulation in the rotating frame (which is justified in the large DC field limit) the value of $\Delta t$ is set by the strongest, nearest neighbor, dipolar coupling, unless the RF field is on, at which point $\Delta t$ is reduced accordingly, as the precession about the RF field is faster than in dipolar fields. To facilitate simulations we have truncated the dipolar Hamiltonian to retain only the nearest neighbor couplings and use cubes with $16^{3}$ spins. Of the various possible finite-size, -range, and -time effects we have found empirically that finite time effects controlled by $\Delta t$ are by far the most serious and focused on minimizing them.  We spot-checked that quoted results are robust against increasing both the interaction range and lattice size.  Lastly, to overcome the low signal-to-noise ratio in such small lattices we artificially reduce the temperature so that $\beta = (k_{B}T)^{-1} = 0.2$ (and we average over 100 initial conditions). Since this value of magnetization is five orders of magnitude higher than that of CaF$_2$ at room temperature in high fields, we explicitly checked that the spin dynamics is still effectively governed by infinite temperature correlations by working at higher temperature (and thus at lower mean magnetization) and only observing enhancement of statistical noise.
\section{\label{sec:results}Results and Discussion}
\subsection{Degradation of the magic echo}
\begin{figure}\begin{center}
\includegraphics[width=3.5in]{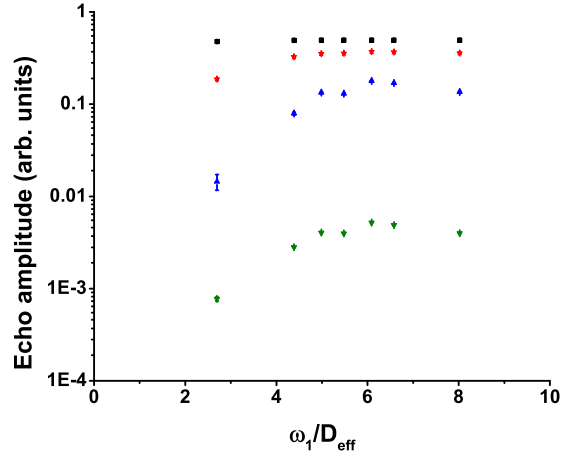}\caption{\label{fig:fpwexp}(Color online) Intensity of the magic echo observed in adamantane for different values of $\tau$ and $\omega_{1}$ corresponding to the RF field
%the $B_1$ field
applied during the $\pi/2$ pulses of the magic sandwich.  As described in the text, the RF amplitude during the spin-locking pulses (periods C and D in Fig. \ref{fig:magicseq}) was kept constant for all experiments.  The values of $\tau$ are $100~\mu$s ($\filledsquare$/black), $200~\mu$s ($\filledstar$/red), $250~\mu$s ($\filledtriangleup$/blue), and $350~\mu$s ($\filledtriangledown$/green) corresponding to $\tau D_{\textrm{eff}} = 4.7, 9.4, 12,$ and $17$ respectively.  As discussed in the text, the variation in the echo amplitude with $\omega_{1}$ and $\tau$ is due largely to the second and higher order terms of the Magnus expansion.}  \end{center}
\end{figure}

\begin{figure}\begin{center}
\includegraphics[width=3.5in]{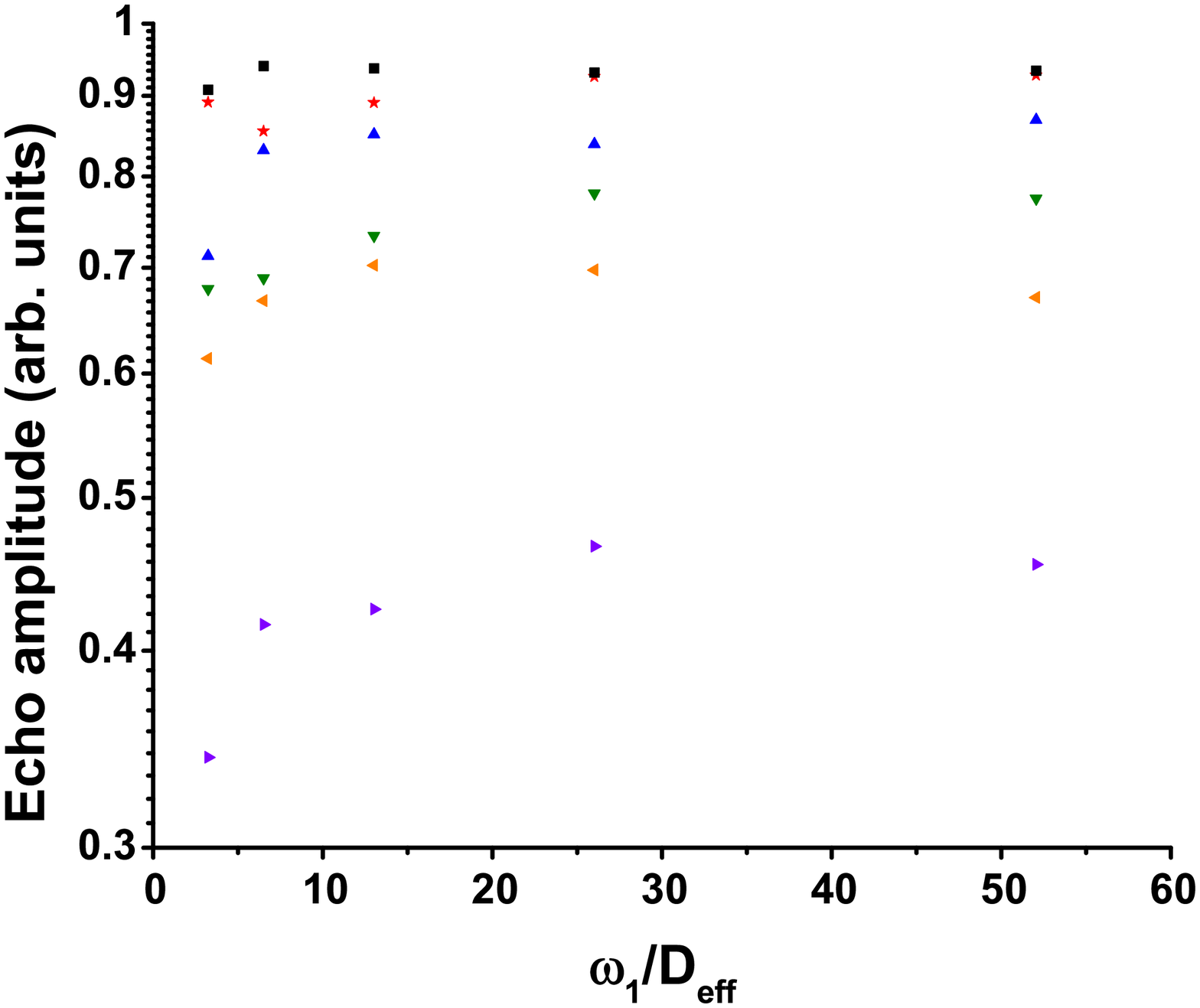}\caption{\label{fig:fpwsim}(Color online) Magic echo amplitudes for lattices of classical spins as a function of $\omega_{1}$
%the $B_1$ field applied
during the $\pi/2$ pulses of the magic sandwich.  As in Fig. \ref{fig:fpwexp} the RF amplitude during the spin-locking pulses (periods C and D in Fig. \ref{fig:magicseq}) was kept constant. The values of $\tau$ are 0.2 ($\filledsquare$/black), 0.5 ($\filledstar$/red), 1 ($\filledtriangleup$/blue), 2 ($\filledtriangledown$/green), 4 ($\filledtriangleleft$/orange), and 8  ($\filledtriangleright$/purple) corresponding  $\tau D_{\textrm{eff}} = 0.14, 0.36, 0.71, 1.4, 2.8,$ and $5.7$, respectively.  The effective dipolar coupling strength is $D_{\textrm{eff}} = 0.71$, as determined by the linewidth.  Note that the values of $\tau D_{\textrm{eff}}$ are smaller here than nearly all the values in Fig. \ref{fig:fpwexp} and that, correspondingly, the variation in echo amplitude is smaller.} \end{center}
\end{figure}

In Figs. \ref{fig:fpwexp} and \ref{fig:fpwsim} we show intensities of the magic echoes in experiments using adamantane and in simulations, respectively, versus  the RF field strength
power applied during the $\pi/2$ pulses $\omega_{1}/D_{\textrm{eff}}$ ($D_{\textrm{eff}}\approx 7 D_{12}$ in a cubic crystal). In the experiments, whose results are shown in Fig. \ref{fig:fpwexp}, the amplitude of the initial $\pi/2$ and all RF bursts was fixed to $8.1 D_{\textrm{eff}}$.  The experiments clearly show a decrease in the echo amplitude with both echo time $\tau$ and with decreasing RF amplitude $\omega_{1}$ of the $\pi/2$ magic sandwich pulses. In Fig. \ref{fig:fpwsim} magic echo amplitudes for varying RF amplitude and $\tau$ are shown for simulations of classical spins.  In these simulations the amplitudes of the RF bursts (during periods C and D) are kept constant at $6.5 D_{\textrm{eff}}$ rads/time.  The data reveal a decrease in the echo amplitude with increasing $\tau$ but the dependence on $\omega_{1}$ is weaker than that observed in the experiments.  These classical simulations became unstable for longer relative echo durations so the quantity $\tau D_{\textrm{eff}}$ was larger in the experiments than in simulations for the largest echo times we used.  When $\tau D_{\textrm{eff}}=5.7$ the variation in the magic echo amplitude decreases by $\sim 21 \%$ when $\omega_{1}/D_{\textrm{eff}}$ is reduced from 6.5 to 3.3. This compares to the experimental case where $\tau D_{\textrm{eff}} = 4.7$ and the echo amplitude decreases by $\sim 4\%$ when $\omega_{1}/D_{\textrm{eff}}$ is reduced from 6.6 to 2.7.  Coupled with the weaker relative RF in the simulations' spin locking periods the classical simulations at least qualitatively capture the features observed in the experiments on adamantane emphasizing the contribution of finite pulse widths in the degradation of the magic echo cycle. Most importantly, the relative influence of finite width $\pi/2$ pulses becomes diminished at \emph{any} $\tau$ with only moderate decrease of $t_{\pi/2}$.
In summary, we experimentally observed a degradation in the magic echo efficacy with decreasing RF amplitude of the magic sandwich $\pi/2$ pulses which becomes pronounced for large echo times.  There is also a flattening-out of the magic echo refocusing for large RF amplitude as increasing the RF amplitude does not provide better refocusing.  While this observation  may be due to instrumental artifacts such as phase transients (which are known to increase with increasing RF amplitude), the fact that it is also borne out in numerical simulations suggests that this flattening is an intrinsic effect, in principle captured by the average Hamiltonian treatment.%, although we are not aware of any such theoretical study (which may require computing yet higher order terms in $\bar{\mathcal{H}}$).
\subsection{Growth, decay, and spread of multispin correlations}
\begin{figure}\begin{center}
\includegraphics[width=3.5in]{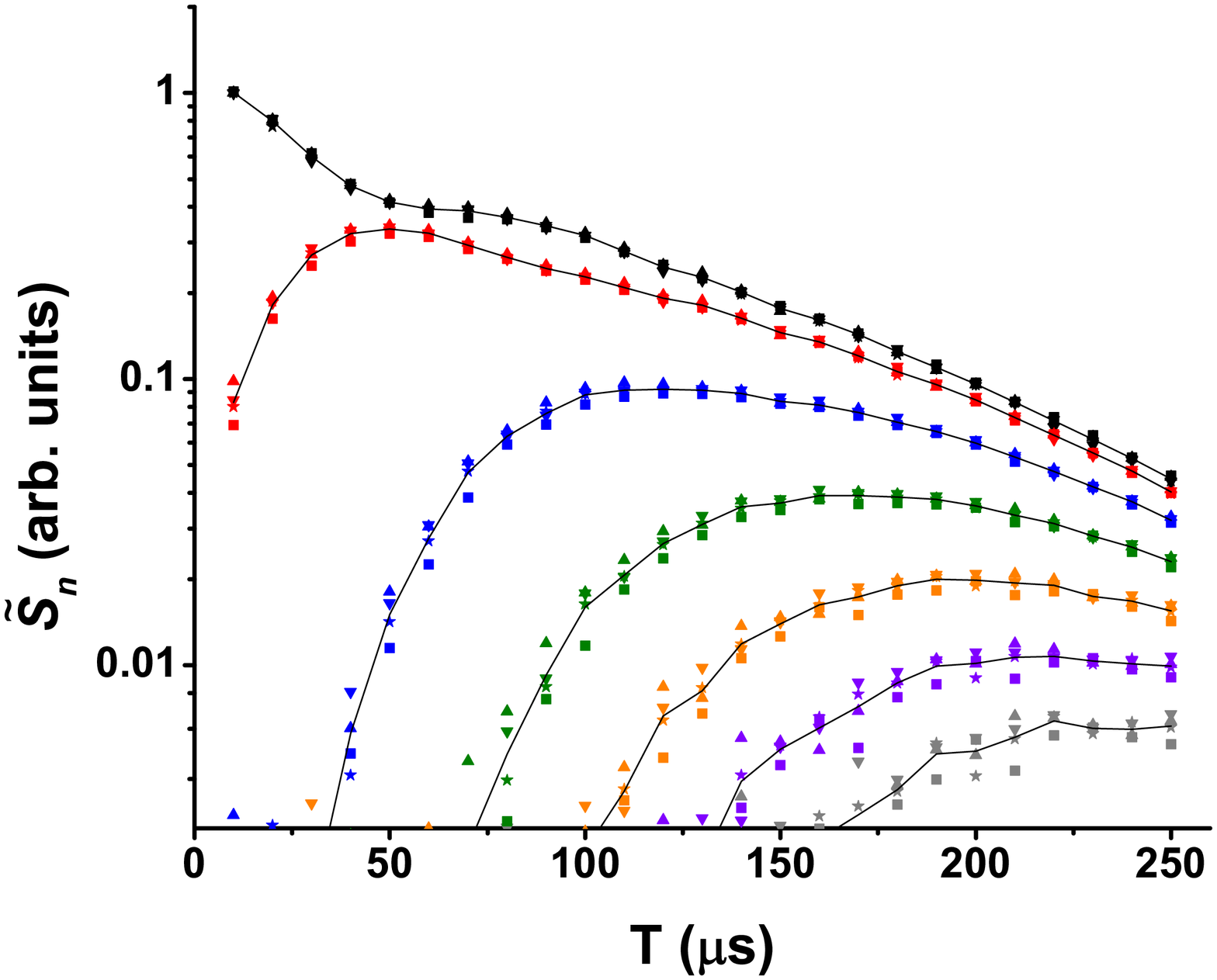}\caption{\label{fig:ltmqc}(Color online) Measured even-order multiple spin correlations in adamantane as a function of time $T$ in Fig. \ref{fig:pulseseq}.  The shapes represent the FID ($\filledsquare$) and echoes at various $\tau$: $\tau$=$100~\mu$s ($\filledstar$), $\tau$=$150~\mu$s ($\filledtriangleup$), and $\tau$= $200~\mu$s ($\filledtriangledown$).  The colors are: zeroth order correlations(black), second order correlations (red), fourth order correlations (blue), sixth order correlations (green), eighth order correlations (orange), tenth order correlations (purple), and are twelfth order correlations (gray).   The signal intensities have been renormalized to emphasize the common long-time decays after $\sim 200~\mu$s. The lines shown are intended to be a guide and do not represent a fit to the experimental data.} \end{center}
\end{figure}

\begin{figure}\begin{center}
\includegraphics[width=3.5in]{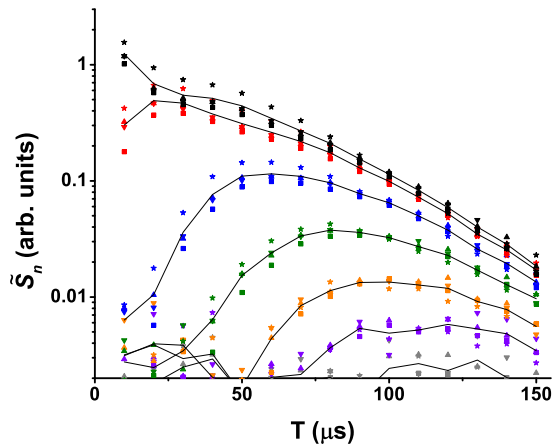}\caption{\label{fig:ltmqcCaF2}(Color online) Measured even-order multiple spin correlations in calcium fluoride as a function of time $T$ in Fig. \ref{fig:pulseseq}.  The shapes represent the FID ($\filledsquare$) and echoes at various $\tau$: $\tau$= $50~\mu$s  ($\filledstar$), $\tau$=$100~\mu$s ($\filledtriangleup$), and $\tau$=$130~\mu$s ($\filledtriangledown$).  The colors are: zeroth order correlations(black), second order correlations (red), fourth order correlations (blue), sixth order correlations (green), eighth order correlations (orange), tenth order correlations (purple), and are twelfth order correlations (gray).  The signal intensities have been renormalized to emphasize the common long-time decays after $\sim 100~\mu$s. The lines shown are intended to be a guide and do not represent a fit to the experimental data.} \end{center}
\end{figure}

In Figs. \ref{fig:ltmqc} and \ref{fig:ltmqcCaF2} we show the various multiple spin correlations plotted as a function of the echo spacing $\tau$ as well as the evolution time $T$ observed in the experiments.  Each echo time's multiple spin correlations in Figs. \ref{fig:ltmqc} and \ref{fig:ltmqcCaF2} have been renormalized separately (one renormalization per echo time $\tau$, for all orders of spin correlations). The adamantane FID data are in good agreement with a previously published study\cite{sanchez2009}.  The agreement between the CaF$_{2}$ data and previously published data \cite{cho2005} is relatively close for short times, although should be noted that in the authors of Ref. \onlinecite{cho2005} normalize the coherence orders so that at every point in time the sum of all orders is one (the correspondence is apparent if one plots the data this way).  However, there are some differences, which become especially apparent in the higher order correlations at longer times.  The orientation of our crystal appears to be different when the FIDs are compared--prior data was acquired with the crystal's [110] axis along the magnetic field \cite{cho2005}.  The initial growth of the multiple spin correlations details a significant difference between the FIDs and the various echoes, revealing imperfections in the time reversal of the magic echo.
We now turn to a comparison of the FID and magic echo data.  Quite generally, higher order correlations appear large in magic echo traces as compared to those of the FID, which is not surprising. It is also apparent that the higher order correlations peak earlier for the echoes than for the FID, an additional indication of imperfect refocusing and leftover multispin correlations at the peaks of the magic echoes which are not present in the FID.

Looking beyond the initial growth stage, Figures \ref{fig:ltmqc} and \ref{fig:ltmqcCaF2} reveal that the multiple spin correlation amplitudes decay at the same rates after $\sim 200~\mu$s for adamantane and $\sim 100~\mu$s for calcium fluoride, despite variations in the initial density matrix.  It has been proposed \cite{fine2005,morgan2008} that in the asymptotic long-time regime where the FIDs decay the same (see Figs. \ref{fig:LTOL},\ref{fig:LTOLCaF2}), all elements of the density matrix share the same eigenvalues and eigenmodes of decay.  Our results clearly show that the decay rates appear to be the same and, equivalently, that the relative amplitudes remain constant. We show the results of the fits of the long-time portions of coherence order $n$ to an exponential decay
\begin{equation}
\label{eq:expdecay}\tilde{S}_n(t) = A_{n}\exp{(-\Gamma_{n}t)}\end{equation} in Table \ref{tb:mqfit} for adamantane and calcium fluoride.  In what follows we give insight into this long time behavior.

\begin{table*}
\caption{\label{tb:mqfit}The fit parameters $\Gamma_{0}$ and $\Gamma_{2}$ for the fit of the zeroth and second order quantum coherence decays with time in the long-time regime in adamantane and calcium fluoride.  The error shown in the table for each value is determined from the standard error of the fit.}
\begin{ruledtabular}
\begin{tabular}{cccccc}
Adamantane & $\Gamma_{0}$ (ms$^{-1}$) & $\Gamma_{2}$ (ms$^{-1}$) & Calcium Fluoride & $\Gamma_{0}$ (ms$^{-1}$) & $\Gamma_{2}$ (ms$^{-1}$) \\ \hline
FID & $13.2 \pm 0.3$ & $13.6 \pm 0.3$ & FID &$35.0 \pm 1.3$ & $33.7 \pm 1.2$  \\
$100~\mu$s magic echo & $ 13.2 \pm 0.3$ & $13.2 \pm 0.5$ & $50~\mu$s magic echo & $36.0 \pm 1.7$ &$35.1 \pm 1.0$\\
$150~\mu$s magic echo & $13.0 \pm 0.4$ & $13.0 \pm 0.4$ & $100~\mu$s magic echo & $34.1 \pm 1.7$ & $34.4 \pm 1.7$\\
$200~\mu$s magic echo & $13.6 \pm 0.4$ & $14.2 \pm 0.3$ & $130~\mu$s magic echo & $34.2 \pm 2.1$ & $32.9 \pm 1.2$\\
\end{tabular}
\end{ruledtabular}
\end{table*}

\subsection{Distribution of multispin correlations and its dynamics}
The coherence order of the time evolved density matrix along the $z$-axis is conserved -- the density matrix only connects states whose total moment's projection onto the $z$-axis differs by $\hbar$ (this is a property of high field, secular dipolar evolution which conserves the total moment along the applied field). The experimentally obtained data resolves build-up and decay of coherence orders \emph{transverse} to the $z$-axis, which reflect %related to growth of
multi-spin correlations in the density matrix. With $m$ and $n$ denoting basis states with different amounts of transverse magnetization, the quantity of interest is
\be
S_\phi=\sum_{m,n} e^{i (n-m) \phi} |\rho_{mn}(t)|^2\equiv\sum_{n=-\infty}^\infty e^{i n \phi} \tilde{S}_n (t).
\ee

Since unitary time evolution preserves the spectrum of the density matrix generally and its purity, $\textrm{Tr}[\rho^2(t)]$,  in particular, there is a convenient sum rule $\sum_n \tilde{S}_n(t)=M^2$, where $M$ is the initial magnetization of the sample.\footnote{Note that in principle there are an infinite number of different sum rules corresponding to higher moments of the density matrix, which may be accessed by adding additional phase encoding steps and performing additional Fourier transforms. We are not aware of any work to understand their significance and for the time being only focus on the one above.}
Also, quite generally,  we expect $\tilde{S}_n(t)$ to be peaked near $n=0$ with its width expanding in time (phenomenologically, it may be approximated as a discrete Gaussian profile, e.g. see Fig. \ref{fig:COHOE}).  Quite trivially, then, the long-time temporal decay of individual $\tilde{S}_n(t)$ is dictated by the growth of the width of the entire profile and the sum rule.  Slightly more formally, we can use the second moment
\be
N_2(t)= \sum_n n^2 \tilde{S}_n(t)
\ee
as a proxy for the distribution's width, and the center of the distribution is expected to decay as $\sim 1/\sqrt{N_2(t)}$.

The basic qualitative aspect of the dynamics is the redistribution of correlations (in the $z$-basis) from few spins to the entire crystal. This can be gleaned from a set of coupled linear differential equations governing (Liouvillian) evolution of multi-spin correlation functions, known for quite some time \cite{lado1971}. Crucially, these equations are completely specified by the moments of the NMR lineshape \cite{vanvleck}. Unfortunately, these moments are not particularly well known beyond a few low orders \cite{jensen1973}, so there is no closed form solution of this problem in the dipolar case of interest here.

However, some progress has been made recently, by Zobov and Lundin \cite{zobov2004,zobov2006}, which we now summarize.
First \cite{zobov2004}, for an infinite range dipolar kernel a wealth of closed form solutions was obtained. For instance,
\be
\tilde{S}_n(t)=I_n(t^2)(1+M_2 n^2/t^2)e^{-M_2 t^2},\ee
 where $M_2$ is the second spectral moment of the NMR lineshape. The long-time behavior of this model shows convergence, albeit non-exponential, of different coherence orders, $\tilde{S}_n\sim 1/t$ for all $n$ as $t\to\infty$ (with $N_2\sim t^2$). While this particular feature is reminiscent of our findings, the infinite range model and many of its properties are not physical although with some modifications it may be used, perhaps, to describe intermediate time behavior.\footnote{The short-time behavior of the infinite range model is quite distinct from any physical system, however these discrepancies can be understood and remedied. This was done to some degree by Ref. \onlinecite{bodneva2009}, which revisited the exact differential equations of Ref. \onlinecite{lado1971} and solved them using an ansatz (which is wrong generally but correct for the infinite range model) whereby all higher moments of the NMR line are assumed to be the same.  In addition to recovering the $\sim 1/t$ behavior above these authors have also documented the sort of syncronization of \emph{phases} of multiple coherent components of the density matrix (our experiment does not measure these). More importantly, their treatment contains the correct, oscillatory short-time behavior of spin-dynamics misrepresented in the infinite range model. Lastly, the infinite range model displays an interesting, almost exponential (very non-Gaussian) dependence on the coherence order. Altogether we suspect that these remarkable features are unphysical, strictly speaking, but may be of relevance to behavior on intermediate times.}

More recently (and realistically), the same group studied the dipolar case \cite{zobov2006} and established simple exponential growth of $N_2(t)\sim \exp(2 C t)$. Our data for adamantane and calcium fluoride supports this prediction qualitatively, though this measurement is difficult, at least partially because of the encoding of finite numbers of spins in our experiments. However, our fits for the long-time portions for calcium fluoride gives a rate constant $C$ on the order of $0.02~\mu$s$^{-1}$ in calcium fluoride and $0.008~\mu$s$^{-1}$ in adamantane.  Combining this result with an assumption of a simple Gaussian profile (empirically justified for our data but likely more complicated in reality), we may expect a simple \emph{exponential} decay of $\tilde{S}_n(t)\sim \exp(-\Gamma_n t)$, with approximately constant $\Gamma_n$, which is also consistent with the data in Figs. \ref{fig:LTOL}, \ref{fig:LTOLCaF2} and in Table \ref{tb:mqfit}.

In connection with predictions based on microscopic chaos \cite{fine2005}, we note that the decay rates $\Gamma_{0}$ and $\Gamma_{2}$ of the zeroth and second order correlations matches the FID decay rate $g$ in CaF$_{2}$ \footnote{We also confirmed the finding that $\Gamma_{0}$, $\Gamma_{2}$, and $g$ for the FID are the same, in calcium fluoride samples with different orientations, using data provided by H. Cho.}.  The same is not true in adamantane, which could be a result of residual motional averaging in adamantane that makes it a less than ideal test case for rigid lattice NMR.  In particular, the agreement of the decay rates of various coherence orders is consistent with the prediction of Ref. \onlinecite{fine2005}, which predicts that the entire density matrix shares common decay properties in the long-time limit.

\section{Summary and outlook}
This work aimed to apply novel multiparticle techniques for measuring the quality of time reversal in a magic echo. In the process, we have provided a new quantitative understanding of realistic magic echoes, but also caught a glimpse of the fundamental and universal process by which closed many-body systems reach \emph{apparent} thermalization or pseudo-thermalization.  Simple athermal correlations (easy to produce and to measure) evolve unitarily into ever more delicate and difficult to observe multiparticle objects. While the decay of the former is commonly observed and equated with thermalization, one of the accomplished goals of the present study was to document the latter process.
%Our , with apparent thermalization (or pseudo-thermalization) taking place once observable elements of the density matrix have decayed.

It would be of great interest to compare and contrast this  physics under different conditions and, especially, with increased signal-to-noise, e.g. using cryprobes\cite{Styles1984}.
%We have observed significant differences, both quantitative and qualitative, between the multiple spin correlations following a magic echo and the FID.  However, in connection with the single-spin decays observed directly in NMR experiments, the long-time portions of the multiple spin correlations converge for the FIDs and for the different magic echoes.  This gives insight into the spin dynamics of this long-time region and should be the subject of further study with increased signal-to-noise from the use of cryoprobes \cite{Styles1984} or other techniques to increase sensitivity.
Of particular fundamental interest is the influence of the environmental degrees of freedom -- the general expectation is that external sources of decoherence in the spin dynamics will cut off the growth process. In this case the decay of low order correlations will be due to extrinsic noise.

\section*{Acknowledgments}
%V. Oganesyan and G. S. Boutis
We would like to acknowledge support from the City University of New York Collaborative Incentive Research Grant Round 16.  This research was also supported, in part, under National Science Foundation Grants CNS0958379 and CNS0855217  and the City University of New York High Performance Computing Center.  G.S. Boutis also acknowledges support from award No. SC1GM086268-04 from the National Institute of General Medical Sciences. The content is solely the responsibility of the authors and does not necessarily represent the official views of the National Institute of General Medical Sciences or the National Institutes of Health (NIH). V. Oganesyan gratefully acknowledges the support of the NSF through award DMR-0955714, as well hospitality of CNRS and Institute Henri Poincar\'{e} (Paris, France) and KITP (Santa Barbara), where this research was supported in part by the National Science Foundation under Grant No. NSF PHY11-25915. 
We also thank S. Barrett, D. Huse, and C. Ramanathan for helpful discussions and comments, and H. Cho for sharing experimental data related to Ref. \onlinecite{cho2005}.

\bibliography{MQME}

\end{document}